\documentclass[12pt,a4paper,final]{iopart}

\usepackage{iopams}  
\usepackage{graphicx}
\usepackage[breaklinks=true,colorlinks=true,linkcolor=blue,urlcolor=blue,citecolor=blue]{hyperref}
\usepackage{breqn}
\usepackage{subfigmat}

\expandafter\let\csname equation*\endcsname\relax
\expandafter\let\csname endequation*\endcsname\relax

\begin{document}

\title{Global fluid simulation of plasma turbulence in a stellarator with an island divertor}

\author{A. J. Coelho, J. Loizu, P. Ricci, M. Giacomin}
\address{\'Ecole Polytechnique Fédérale de Lausanne (EPFL), Swiss Plasma Center (SPC), CH-1015 Lausanne, Switzerland}
\ead{antonio.coelho@epfl.ch}

\begin{abstract}
Results of a three-dimensional, flux-driven, electrostatic, global, two-fluid turbulence simulation for a 5-field period stellarator with an island divertor are presented. The numerical simulation is carried out with the GBS code, recently extended to simulate plasma turbulence in non-axisymmetric magnetic equilibria. The vacuum magnetic field used in the simulation is generated with the theory of Dommaschk potentials, and describes a configuration with a central region of nested flux surfaces, surrounded by a chain of magnetic islands, similarly to the diverted configurations of W7-X. The heat outflowing from the core reaches the island region and is transported along the magnetic islands, striking the vessel walls, which correspond to the boundary of the simulation domain.

The radial transport of particles and heat is found to be mainly driven by a field-aligned coherent mode with poloidal number $m=4$. The analysis of this mode, based on non-local linear theory considerations, shows its ballooning nature. In contrast to tokamak simulations and experiments, where blobs often contribute to transport, we do not observe the presence of intermittent transport events. 

\end{abstract}


\newpage

\section{Introduction}

In the last years, the magnetic fusion community has increasingly focused its attention on how particles and heat can be exhausted without damaging the plasma-facing materials, while simultaneously preserving good core performance. Most tokamaks are currently operated in an axisymmetric single-null diverted configuration, consisting of a central region of closed field lines, surrounded by the scrape-off layer (SOL), a region of open field lines that are diverted by using external coils and end up on the solid surfaces of the divertor plates at a certain distance from the core. The single-null configuration will be adopted by ITER~\cite{iter_configuration}, while alternative configurations such as the snowflake or the double-null are being explored for DEMO~\cite{holger_alternative_divertors}. As for stellarators, diverted solutions also exist. In W7-X the so-called \textit{island divertor} configuration is used~\cite{W7X_first_results}, where a chain of magnetic islands surrounds the closed field-line region, and the heat outflowing from the core is diverted along the field lines of the islands that strike the plasma-facing components. Other possible exhaust solutions used in stellarators include the ergodic divertor employed in LHD~\cite{LHD_divertor} and the non-resonant divertor envisaged for HSX~\cite{non_resonant_divertors}.

The interplay between the plasma fluxes from the core, the cross-field transport across the magnetic field lines and the losses to the walls determine the peak heat loads at the vessel targets~\cite{paolo_rogers_2013}. While boundary turbulence has been thoroughly characterized experimentally in tokamaks~\cite{Zweben_edge}, similar investigations are still in their infancy in stellarators. In addition to the measurements carried out in small, low-temperature plasma stellarators such as TJ-K~\cite{TJK_PRL_2009,TJK_PRL_2011,TJK_blobs_2016}, experimental measurements in the edge of W7-X are recent~\cite{quasi_coherent_W7X,filaments_W7X,edge_W7X}. These have shown significant differences with respect to tokamaks: field-aligned structures (dubbed filaments) are observed to be essentially bound to their flux surface, covering very small radial distances due to their slow radial velocity of the order of 100m/s~\cite{quasi_coherent_W7X,filaments_W7X}, while their poloidal velocities are of the order of a few km/s~\cite{edge_W7X}. Similar observations are reported in the edge of LHD~\cite{LHD_edge_1,LHD_edge_2}. On the other hand, typical tokamak measurements show blobs generated near the separatrix that travel balistically across the SOL and provide an important contribution to the global perpendicular transport, with radial velocities of the order of a few km/s~\cite{cedric_blobs,szweben_blobs}. In W7-X, the fluctuations are approximately normally distributed~\cite{edge_W7X}, hinting that the turbulent structures are the result of fluctuations that have a local origin rather than being advected from a different radial position, in agreement with their small radial velocity~\cite{filaments_W7X}. While a broad-band turbulent spectrum in the range 1-10kHz is observed in stellarators as in tokamaks~\cite{quasi_coherent_W7X,edge_W7X}, quasi-coherent modes with 10-50kHz and small poloidal wave-numbers (corresponding to a wavelength of approximately 30cm), propagating both in the ion and electron diamagnetic drift directions, are also observed in W7-X~\cite{quasi_coherent_W7X}.

Similarly to the experimental investigations, the simulation of plasma dynamics in the stellarator boundary is still at a very early stage. The fluid code BOUT++ simulated seeded plasma filaments in a slab geometry that emulates the radially varying connection lengths of W7-X~\cite{shanahan_jpp_2020} and it was recently extended to simulate non-axisymmetric magnetic fields such as a low-field-period rotating ellipse~\cite{BSTING}. 

In this Letter we present results of the first global, two-fluid, flux-driven simulation of a stellarator with an island divertor. 
The simulation is performed with the GBS code, which has been used in the past decade to simulate the tokamak boundary~\cite{paolo_GBS,giacomin_jcp,maurizio_snowflake,carrie_paper,maurizio_turbulent_regimes,mancini_paper,maurizio_scalings}, and it is here extended to non-axisymmetric magnetic fields. GBS solves the drift-reduced Braginskii equations~\cite{zeiler}, valid in the high collisionality regime, which is often justified in the plasma boundary of magnetic fusion devices as well as in the core of low-temperature devices (e.g., TJ-K \cite{TJ-K}). 
All quantities are evolved in time without separation between equilibrium and fluctuating parts. We consider here the electrostatic limit, we apply the Boussinesq approximation~\cite{paolo_GBS}, and we neglect gyroviscous terms as well as the coupling to the neutral dynamics, although these are implemented in the most recent version of the GBS code for tokamak simulations~\cite{giacomin_jcp}. Within these approximations, the drift-reduced model evolved by GBS is:

\begin{equation}
    \pt{n} = -\frac{\rorhos}{B}\left[\Phi,n\right] - \nablapar(n\vpare) + \frac{2}{B}\left[C(p_e)-nC(\Phi)\right] + D_n\nablaperp^2n + D_n^{\parallel}\nablapar^2n +  \mathcal{S}_n 
    \label{eq:density}
\end{equation}

\begin{equation}
\begin{aligned}
    \pt{T_e} = &-\frac{\rorhos}{B}\left[\Phi,T_e\right] - \vpare\nablapar T_e + \frac{4T_e}{3B}\left[\frac{C(p_e)}{n}+\frac{5}{2}C(T_e)-C(\Phi)\right]\\ &+ \frac{2T_e}{3n}\left[0.71\nablapar j_{\parallel}-n\nablapar\vpare\right] + D_{T_e}\nablaperp^2T_e+\chi_{\parallel e}\nablapar^2T_e + \mathcal{S}_{T_e}
\end{aligned}
\end{equation}

\begin{equation}
\begin{aligned}
    \pt{T_i} = &-\frac{\rorhos}{B}\left[\Phi,T_i\right] - \vpari\nablapar T_i + \frac{4T_i}{3B}\left[\frac{C(p_e)}{n}-\frac{5}{2}\tau C(T_i)-C(\Phi)\right]\\ &+ \frac{2T_i}{3n}\left[\nablapar j_{\parallel} - n\nablapar\vpari\right] + D_{T_i}\nablaperp^2T_i + \chi_{\parallel i}\nablapar^2T_i + \mathcal{S}_{T_i}
\end{aligned}
\end{equation}

\begin{equation}
\begin{aligned}
    \pt{\vpare} = &-\frac{\rorhos}{B}\left[\Phi,\vpare\right]-\vpare\nablapar\vpare + \frac{m_i}{m_e}\left[\nu j_{\parallel}+\nablapar\Phi-\frac{\nablapar p_e}{n} - 0.71\nablapar T_e\right]\\ &+ \eta_{0e}\nablapar^2\vpare + D_{\vpare}\nablaperp^2\vpare
    \label{eq:vpare}
\end{aligned}
\end{equation}

\begin{equation}
    \pt{\vpari} = -\frac{\rorhos}{B}\left[\Phi,\vpari\right]-\vpari\nablapar\vpari - \frac{1}{n}\nablapar(p_e+\tau p_i)+\eta_{0i}\nablapar^2\vpari + D_{\vpari}\nablaperp^2\vpari
\end{equation}

\begin{equation}
    \pt{\omega} = -\frac{\rorhos}{B}\left[\Phi,\omega\right] -\vpari\nablapar\omega + \frac{B^2}{n}\nablapar j_{\parallel} + \frac{2B}{n}C(p_e+\tau p_i) + D_{\omega}\nablaperp^2\omega + D_{\omega}^{\parallel}\nablapar^2\omega
    \label{eq:vorticity}
\end{equation}

\begin{equation}
    \nablaperp^2\Phi = \omega - \tau\nablaperp^2T_i
    \label{eq:potential}
\end{equation}
In Eqs. (\ref{eq:density})-(\ref{eq:potential}), all quantities are normalized to reference values: density $n$, electron temperature $T_e$ and ion temperature $T_i$ are normalized to $n_0$, $T_{e0}$ and $T_{i0}$; electron parallel velocity $\vpare$ and ion parallel velocity $\vpari$ are both normalized to the sound speed $c_{s0}=\sqrt{T_{e0}/m_i}$; vorticity $\omega$ and the electrostatic potential $\Phi$ are normalized to $T_{e0}/(e\rhos^2)$ and $T_{e0}/e$; time is normalized to $R_0/c_{s0}$, where $R_0$ is the machine major radius; and perpendicular and parallel lengths are normalized to the ion sonic Larmor radius, $\rhos=\sqrt{T_{e0}m_i}/(eB_0)$, and $R_0$, respectively. The normalized parallel current is $j_{\parallel}=n(\vpari-\vpare)$ and $B$ is the magnetic field normalized to the magnitude of the field on axis, $B_0$.

The dimensionless parameters appearing in the equations are the normalized ion sonic Larmor radius $\rho_*=\rhos/R_0$, the normalized electron and ion parallel diffusivities, $\chi_{\parallel e}$ and $\chi_{\parallel i}$, considered here as constants, the ion to electron temperature ratio $\tau=T_{i0}/T_{e0}$, the normalized electron and ion viscosities, $\eta_{0e}$ and $\eta_{0i}$, which we also set to constant values, and the normalized Spitzer resistivity $\nu=\nu_0T_e^{3/2}$ with $\nu_0$ given in \cite{maurizio_turbulent_regimes}. Small numerical diffusion terms such as $D_n\nablaperp^2n$ and $D_n^{\parallel}\nablapar^2n$ (and similar for the other fields) are introduced to improve the numerical stability of the code and the simulation results show that they lead to significantly lower perpendicular transport than turbulence. The terms $\mathcal{S}_n$, $\mathcal{S}_{T_e}$ and $\mathcal{S}_{T_i}$ denote the sources of density, electron and ion temperature, respectively. Magnetic pre-sheath boundary conditions, described in \cite{joaquim_BCs,annamaria_BCs}, are applied to all quantities at the end of the field lines intersecting the walls, except for the density and vorticity, that satisfy, respectively, $\partial_sn=0$ and $\omega=0$, where $s$ is the direction normal to the wall. 

The normalized geometrical operators appearing in Eqs. (\ref{eq:density})-(\ref{eq:potential}) are the parallel gradient $\nabla_{\parallel}u = \boldsymbol{b}\cdot\boldsymbol{\nabla}u$, the Poisson brackets $[\Phi,u]=\boldsymbol{b}\cdot\left[\boldsymbol{\nabla}\Phi\times\boldsymbol{\nabla} u\right]$, the curvature operator $C(u) = (B/2)\left[\boldsymbol{\nabla}\times(\boldsymbol{b}/B)\right]\cdot\boldsymbol{\nabla}u$, the parallel Laplacian $ \nabla_{\parallel}^2u = \boldsymbol{b}\cdot\boldsymbol{\nabla}(\boldsymbol{b}\cdot\boldsymbol{\nabla}u)$ and the perpendicular Laplacian $\nabla_{\bot}^2u = \boldsymbol{\nabla}\cdot\left[(\boldsymbol{b}\times\boldsymbol{\nabla}u)\times\boldsymbol{b}\right]$. In contrast to tokamak simulations, these operators are not toroidally symmetric. For their numerical implementation, we expand them in the following small parameters: the ratio between the poloidal components and the norm of the magnetic field, $\delta=B_R/B\sim B_Z/B$; the normalized mirror ratio, $\Delta=(B_{\text{max}}-B_{\text{min}})/\overline{B}$, where $\overline{B}$ is the averaged value of $B$ along the toroidal direction and $(B_{\text{max}}-B_{\text{min}})$ is the ripple amplitude; and the ratio between perpendicular and parallel turbulence length scales, $\sigma=\ls/\lp$. We then retain only the leading order terms in these expansion parameters. 
For the stellarator configuration considered in this work $\delta\sim0.1$, $\Delta\lesssim0.1$ and \textit{a posteriori} we verify that $\sigma\sim 0.01$, confirming the validity of the expansion.

The physical model in Eqs.~(\ref{eq:density})-(\ref{eq:potential}) is discretized in a cylindrical grid $(R,\phi,Z)$, with $R$ the radial coordinate, $\phi$ the toroidal angle and $Z$ the vertical coordinate. The simulation domain is a torus of radius $R_0$ with a rectangular cross-section of size $L_R\times L_Z$. Equations (\ref{eq:density})-(\ref{eq:vorticity}) are advanced in time with an explicit Runge-Kutta fourth-order scheme, while spatial derivatives are computed with a fourth-order finite difference scheme~\cite{paruta_paper,maurizio_turbulent_regimes}.

\begin{figure}[]
 \centering
 \includegraphics[width=16cm]{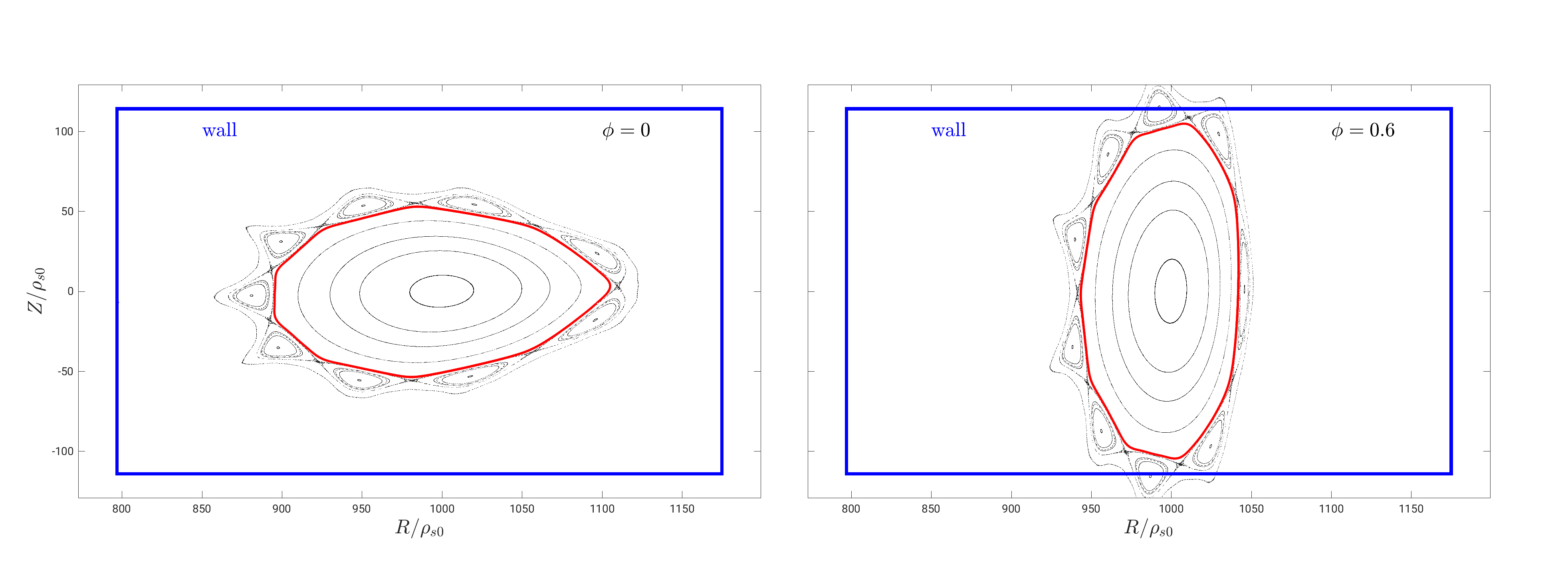}
 \caption{Poincar\'e plot of the magnetic field used in this work on two poloidal planes at different toroidal angles. The boundary of the simulation domain is represented in blue.}
 \label{fig:poincare}
\end{figure}

For the present study we exploit the properties of Dommasck potentials~\cite{dommashck_1986} to analytically construct a five-field period stellarator-symmetric vacuum field with a 5/9 chain of islands surrounding a volume of nested magnetic surfaces, which consist essentially of rotating ellipses without torsion. The magnetic shear is very small with the rotational transform varying from 0.500 at the magnetic axis and 0.555 at the last closed flux surface (LCFS). Since the LCFS is not well defined in a stellarator, we consider it as being a surface very close to the island chain (red surface in Fig.1.). The lengths $L_R$ and $L_Z$ are adjusted such that the islands strike the top and bottom of the simulation box, as can be seen in Fig.~\ref{fig:poincare}. In such configuration, heat and particles outflowing from the core reach the island region and are transported along the magnetic field of the islands, eventually striking the top and bottom walls at specific toroidal locations.


\begin{figure}[]
 \centering
 \includegraphics[width=11cm]{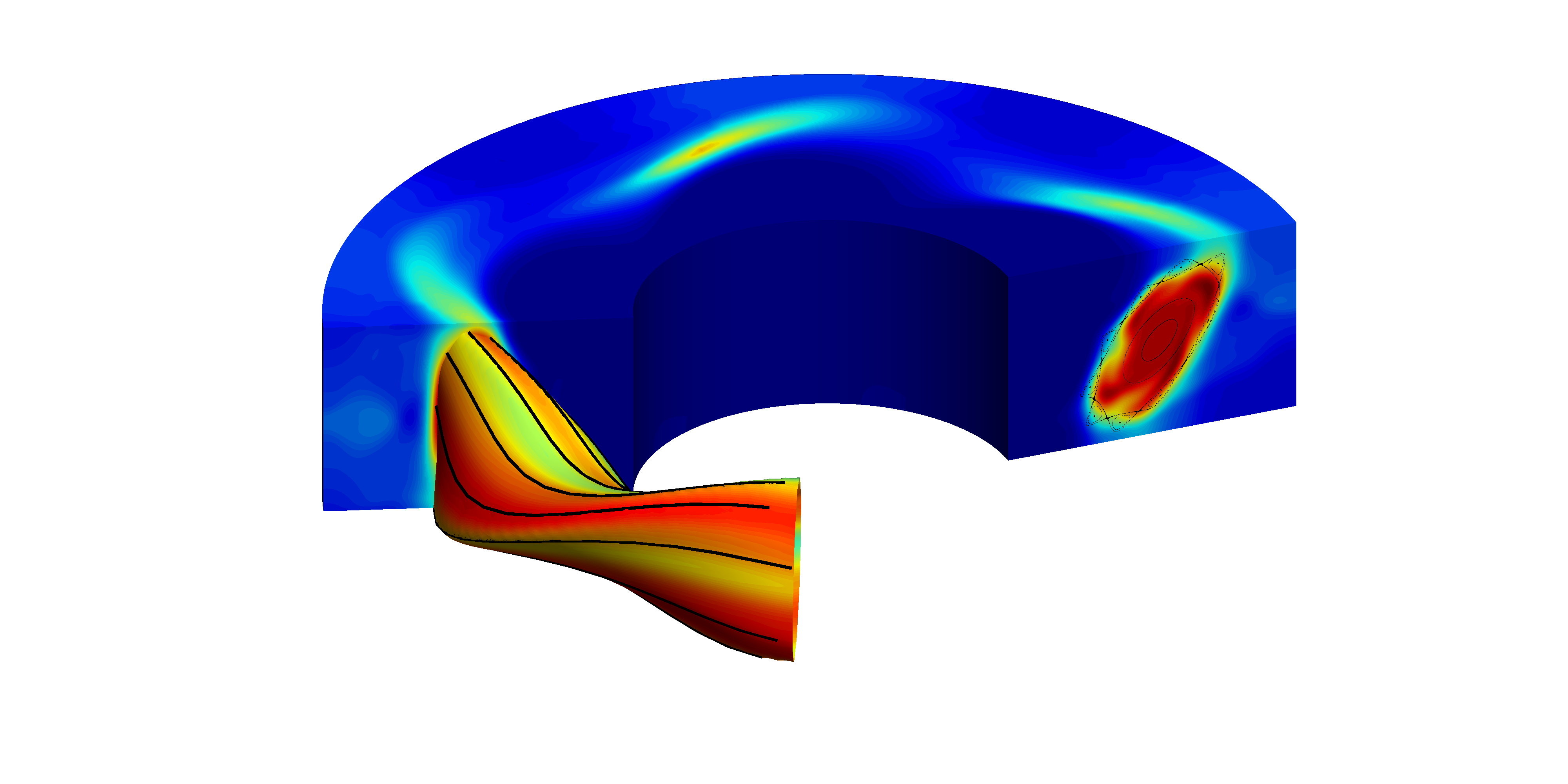}
 \caption{Three-dimensional snapshot of plasma density, representing the geometry of the simulation considered here. The surface depicted corresponds to the LCFS and a field line (black line) is superimposed.}
 \label{fig:gbs_simul_glob}
\end{figure}

The simulation presented here is started from a noisy initial state and, after a transient, reaches a quasi-steady state, where sources, parallel and perpendicular transport and losses at vessel balance each other. We focus our discussion on the results of a simulation that uses the following parameters: $\rorhos=1000$, $\nu_0=0.1$, $\tau=1$, $\chi_{\parallel e,i}=\eta_{0 e,i}=1.0$, $D_n=D_{Te}=D_{Ti}=D_{\vpare}=D_{\vpari}=D_{\omega}=10$, $D_{n}^{\parallel}=D_{\omega}^{\parallel}=1$, $L_R=380\rhos$, $L_Z=230\rhos$, a grid resolution of $N_R\times N_Z\times N_{\phi}=200\times 120\times200$ points and a time-step of $2.9\times 10^{-6}R_0/c_{s0}$. The sources for density and temperature, $\mathcal{S}_n=\mathcal{S}_{T_e}=\mathcal{S}_{T_i}$, are localized around a magnetic surface near the LCFS. 
In Fig.~\ref{fig:gbs_simul_glob}, a global overview of the simulation geometry is shown with a snapshot of density once the quasi-steady state is reached. Particles are mostly exhausted where field lines strike the vessel, which is revealed by the high plasma density regions appearing on the top wall.

\begin{figure}[]
 \centering
 \includegraphics[width=17cm]{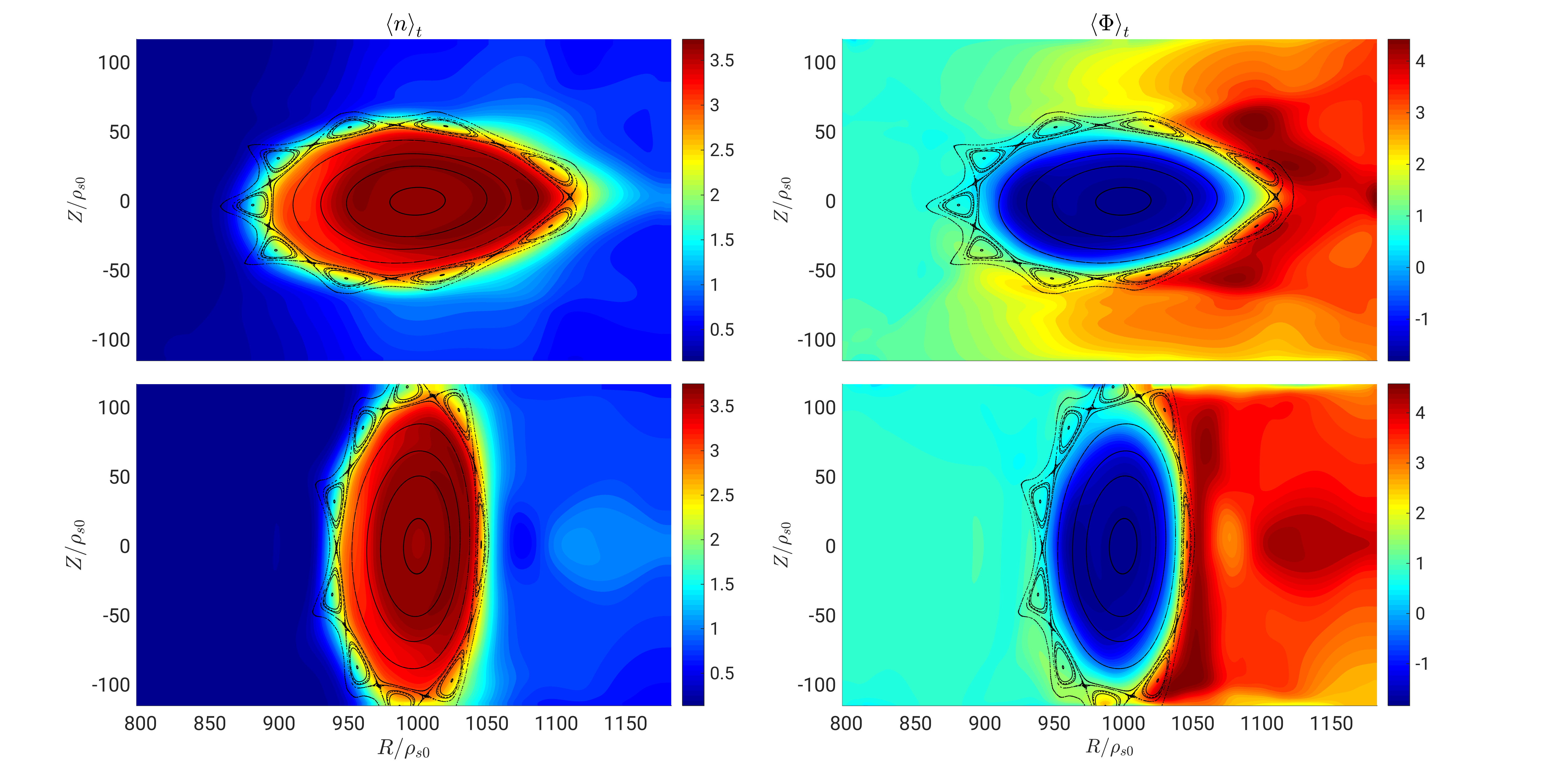}
 \caption{Equilibrium profiles of density (left) and electrostatic potential (right), obtained by time-averaging the simulation results. Top and bottom correspond to the toroidal planes $\phi=0$ and $\phi=0.6$, respectively.}
 \label{fig:equilibrium_plots}
\end{figure}

In Fig.~\ref{fig:equilibrium_plots} the equilibrium (i.e., time-averaged) profiles of density and potential are shown at two different toroidal locations. We observe that the plasma is well confined inside the LCFS, and that the electrostatic potential is negative in the core and positive in the edge. The radial electric field is therefore negative, which corresponds to the ion-root regime, expected as well in the neoclassical high-collisional regime~\cite{ion_root_per}.

\begin{figure}[]
 \centering
 \includegraphics[width=17cm]{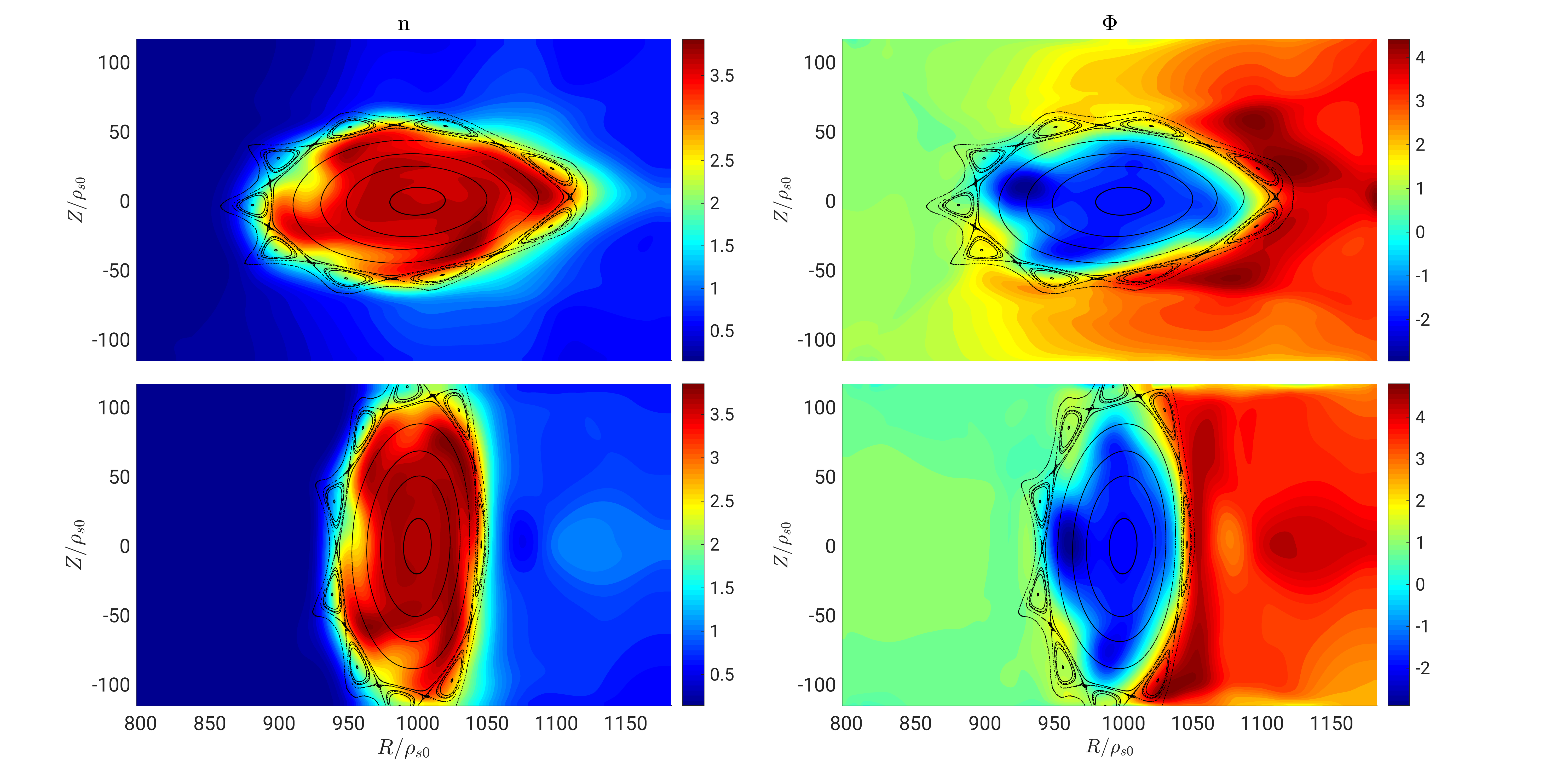}
 \caption{Snapshot of density, $n$ (left) and electrostatic potential, $\Phi$ (right). Top and bottom correspond to the toroidal planes $\phi=0$ and $\phi=0.6$, respectively.}
 \label{fig:mode_plots}
\end{figure}

Snapshots of density and electrostatic potential in the quasi-steady state at two different toroidal angles are shown in Fig.~\ref{fig:mode_plots}. A mode with poloidal mode number $m=4$ (corresponding to $k_y\rhos\approx 0.04$, where $y$ is the binormal direction, being $x$ the radial and $z$ the parallel coordinates) and toroidal mode number $n=5$ dominates the global dynamics of the system. The mode is coherent and rotates in the ion diamagnetic direction. Furthermore, the mode is field-aligned, as can be seen in Fig.~\ref{fig:gbs_simul_glob}, where a field line has been traced for several toroidal transits on the LCFS and is shown with a black line. No broad-band turbulence nor blobs are observed, highly contrasting with typical tokamak boundary simulations~\cite{paola_blobs,shanahan_ppcf_2016}. The presence of a coherent dominant mode is a robust feature of the simulation. In fact, an $m=4$ mode appears in other simulations performed with lower and higher strength of the plasma and temperature sources, and also in simulations with larger sizes, $\rorhos=1500, 2000$ (not shown).

\begin{figure}[]
 \centering
 \includegraphics[width=15cm]{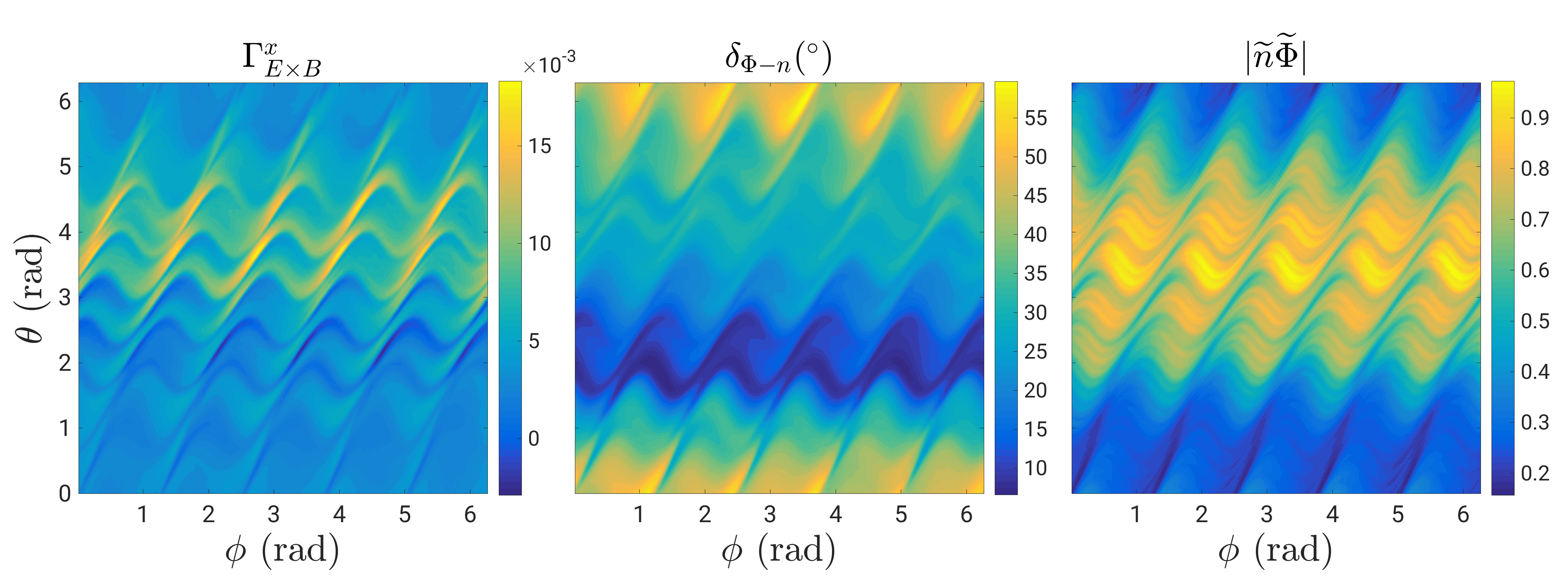}
 \caption{Time-averaged fluctuating $E\PLH B$ radial flux on the LCFS evaluated according to Eq.~(\ref{eq:ExB_flux}) (left panel); phase-difference between potential and density, $\delta_{\Phi-n}$ (middle panel); fluctuations' amplitude (right panel); as evaluated from the GBS simulation. $\theta=0$ corresponds to the outboard midplane.}
 \label{fig:flux_unfolded_plots}
\end{figure}

An analysis on the balance of the density equation shows that approximately 80\% of the radial transport of particles across the LCFS is due to the time-averaged fluctuating $E\PLH B$-flux, 

\begin{equation}
    {\Gamma}^x_{E\times B}=\left<\widetilde{n}\widetilde{V}^x_{E\times B}\right>_t=-\left<\frac{\widetilde{n}}{B}\left(\nabla\widetilde{\Phi}\times\mathbf{b}\right)_x\right>_t,
    \label{eq:ExB_flux}
\end{equation}
where we denote with tilde fluctuating quantities and $\left<\cdot\right>_t$ time averages (i.e., $n=\tilde{n}+\left<n\right>_t$). The largest contribution to this flux is given by the $m=4$ mode. Indeed, we have verified a good agreement between the flux of particles evaluated numerically by using Eq.~(\ref{eq:ExB_flux}), and the time-averaged $E\PLH B$-flux due to a single coherent mode,
\begin{equation}
    \Gamma^c=\frac{k_y}{2B}|\widetilde{n}\widetilde{\phi}|\sin(\delta_{\Phi-n}),
    \label{eq:flux_coherent_mode}
\end{equation}
where $\delta_{\Phi-n}$ is the phase-difference between potential and density obtained from the correlation between them, being the fluctuation amplitudes and phase-difference from the simulation results. In Fig.~\ref{fig:flux_unfolded_plots} we show the $E\PLH B$-flux evaluated using Eq. (\ref{eq:ExB_flux}) on the unfolded LCFS (left panel of Fig.~\ref{fig:flux_unfolded_plots}), where it is seen that the flux peaks on the stellarator high-field side ($\theta=0$ corresponding to the outboard midplane). This is due to the fact that the amplitude of the fluctuations is larger at $\theta=\pi$ (right panel of Fig.~\ref{fig:flux_unfolded_plots}), overcoming the phase-difference term, which is larger on the low-field side (middle panel of Fig.~\ref{fig:flux_unfolded_plots}). 




\begin{figure}[]
 \centering
 \includegraphics[width=10cm]{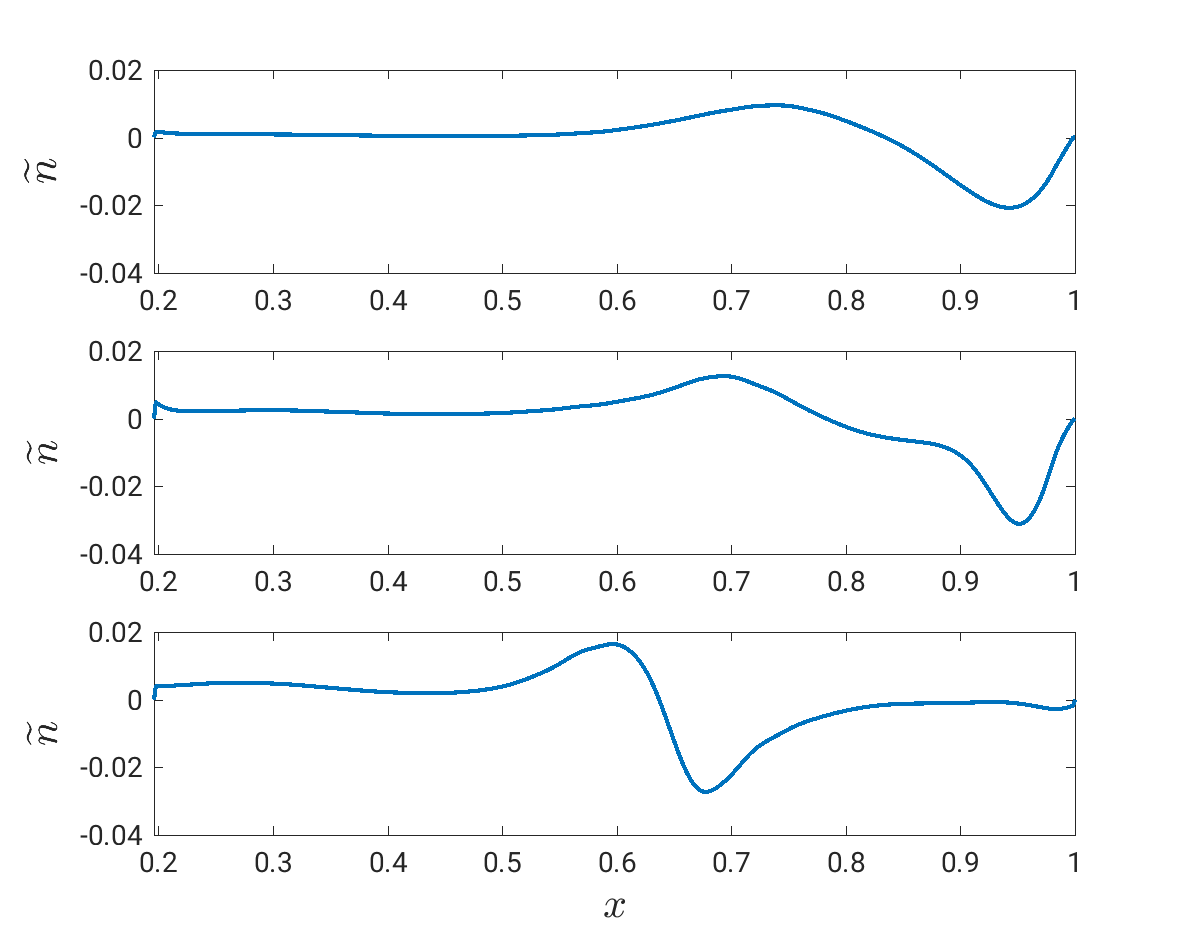}
 \caption{Density eigenmodes resulting from the linear theory with $m=4$ and $n=5$: including all driving terms (top panel); excluding the drift-waves drive (middle panel); excluding the ballooning drive (bottom panel).}
 \label{fig:eigenmodes}
\end{figure}

We use a non-local linear theory in order to investigate the properties of the coherent mode dominating the simulation. Namely, we linearize GBS equations, Eqs. (\ref{eq:density})-(\ref{eq:potential}), by assuming that all quantities vary as
\begin{equation}
    u(x,\theta,\phi) = u_0(x) + \tilde{u}(x)e^{\gamma t+i\left(m\theta-n\phi\right)},
\end{equation}
where $x$ ranges from $0$ at the magnetic axis to 1 at the LCFS, $u_0$ is the background profile obtained from the GBS simulation by time and flux surface averaging, $m$ and $n$ are the poloidal and toroidal mode numbers and $\gamma$ is the growth-rate of the mode. Note that the mode $\widetilde{u}$ depends on $x$, in agreement with the non-local nature of the mode dominating the simulation. In the linearized system of equations, we simplify the magnetic curvature operator by assuming a circular tokamak geometry, motivated by the fact that in the stellarator simulated in this work, the magnitude of the magnetic field varies, up to first order, with $1/R$. After linearizing the equations and evaluating the curvature operator at the low-field side, the study of the linear modes reduces to an eigenvalue equation. In Fig.~\ref{fig:eigenmodes} (top panel), we  show the eigenmode solution with $m=4$ and $n=5$, corresponding to the wavenumbers of the coherent mode observed in the nonlinear simulation. Similarly to the perturbation appearing in the the non-linear simulation, the linear mode is large close to the LCFS and its amplitude decreases towards the core. 

We now show that the identified linear mode is able to transport the radial $E\PLH B$-flux of the non-linear simulation. For this purpose, we consider the $E\PLH B$-flux as given by Eq.~(\ref{eq:flux_coherent_mode}) for the linear eigenmode. The phase difference is obtained simply by computing the phase between the density and potential eigenfunctions of the mode. We determine the amplitude of the linear mode from a balance between the radial $E\PLH B$-flux and the plasma source:

\begin{equation}
    \int_{\partial\Omega}\Gamma^c dS = \int_{\Omega}\mathcal{S}_n\rho_{*} dV,
    \label{eq:balance}
\end{equation}
where $\Omega$ is the volume contained inside the LCFS and $\partial\Omega$ its surface area. Using Eqs.~(\ref{eq:flux_coherent_mode}) and (\ref{eq:balance}) we obtain

\begin{equation}
    |\widetilde{n}\widetilde{\phi}|\sim\frac{2B}{k_y}\frac{\int_{\Omega}\mathcal{S}_n\rho_* dV}{\int_{\partial\Omega}dS}\frac{1}{\sin(\delta_{\Phi-n})}.
    \label{eq:n0phi0}
\end{equation}
While for most ($m,n$) modes the amplitude of the perturbations obtained from Eq.~(\ref{eq:n0phi0}) exceeds by orders of magnitude the one observed in the non-linear simulation, for the $m=4$, $n=5$ mode we obtain $|\widetilde{n}\widetilde{\phi}|\sim0.4$, which is comparable to the values observed in the non-linear simulation (see Fig.~\ref{fig:flux_unfolded_plots}).  


Finally, we address the nature of the $m=4$ mode by removing either the ballooning or the drift-waves instabilities from the linear equations. The drive of the former is removed by zeroing out the curvature term of the vorticity equation, Eq.~(\ref{eq:vorticity}). On the other hand, by removing the parallel gradient terms of temperature and pressure in Ohm's law, Eq.~(\ref{eq:vpare}), we preclude drift-waves from the system. As shown in Fig.~\ref{fig:eigenmodes}, the eigenmode is not significantly affected by the removal of the drift-waves (middle panel). On the other hand, the eigenmode without ballooning drive (bottom panel) is significantly different and is no longer localized in the edge region. Thus, the linear theory points to a ballooning-like nature of the mode that dominates the simulation, a finding to be confirmed by linear analysis that take into account a more complex geometry.

To conclude, this Letter presents the first global fluid simulation of the plasma dynamics in a stellarator with an island divertor. We show that, in the particular configuration produced with the Dommashck potentials considered here, the dynamics of the plasma is dominated by a coherent low ($m=4$) poloidal mode, with no blobs detaching into the SOL. The mode is studied with a non-local linear model, which points out the ballooning-like nature of this mode. These results highly contrast with typical tokamak boundary simulations and investigations are ongoing to understand the cause of this difference, as well as the link with experimental observations.

\ack

The authors thank Per Helander and Brendan Shannahan for useful discussions. This work, supported in part by the Swiss National Science Foundation, has been carried out within the framework of the EUROfusion Consortium and has received funding from the Euratom research and training programme 2014-2018 and 2019-2020 under grant agreement No 633053. The views and opinions expressed herein do not necessarily reflect those of the European Commission.

\section*{References}
\bibliographystyle{unsrt}
\bibliography{bibliography}

\newpage

\end{document}